\documentclass[11pt]{amsart}
\usepackage{geometry}                
\usepackage{graphicx}
\usepackage{amssymb}
\usepackage{amsrefs}
\usepackage{enotez}
\usepackage{epstopdf}
\usepackage[OT2,T1]{fontenc}
\DeclareSymbolFont{cyrletters}{OT2}{wncyr}{m}{n}
\DeclareMathSymbol{\Sha}{\mathalpha}{cyrletters}{"58}

\title[Sub specie aeternitatis: Fourier Transforms]{Sub Specie Aeternitatis: Fourier Transforms from the Theory of Heat to Musical Signals}
\author{Victor Lazzarini}
\date{}                                           

\begin{document}
\maketitle

\begin{abstract}
J. B. Fourier in his \emph{Th\'{e}orie Analytique de la Chaleur}  of 1822 introduced, amongst other things, two ideas
that have made a fundamental impact in fields as diverse as Mathematical Physics, Electrical Engineering,
Computer Science, and Music. The first one of these, a method to find the coefficients for a trigonometric
series describing an arbitrary function, was very early on picked up by G. Ohm and H. Helmholtz as the foundation
for a theory of \emph{musical tones}. The second one, which is described by Fourier's double integral, became
the basis for treating certain kinds of infinity in discontinuous functions, as shown by A. De Morgan in his 
1842 \emph{The Differential and Integral Calculus}. Both make up the fundamental basis for what is now commonly known as the
\emph{Fourier theorem}. With the help of P. A. M. Dirac's insights into the nature of these infinities, we can
have a compact description of the frequency spectrum of a function of time, or conversely of a
waveform corresponding to a given function of frequency. This paper, using solely primary sources, takes
us from the physics of heat propagation to the modern theory of musical signals. It concludes with some 
considerations on the inherent duality of time and frequency emerging from Fourier's theorem.
\end{abstract}

\section*{Dramatis Personae}

\begin{center}
\vspace{0.4cm}
$X(f) = \int_{-\infty}^{\infty} x(t)e^{-i2\pi f t } dt$.............................................................the Fourier transform,

\vspace{0.4cm}
$x(t) = \int_{-\infty}^{\infty} X(f)e^{i2\pi ft} df$..........................................................................and its inverse,

\vspace{0.4cm}
$x(t) = \sum_{k=-\infty}^{\infty} X(kF)e^{i2\pi kFt}$...............................................................the Fourier series,

\vspace{0.4cm}
$X(f) = \sum_{n=-\infty}^{\infty} x(nT)e^{-i2\pi fnT}$...............................the Discrete-time Fourier transform,

\vspace{0.4cm}
$X(k) = \sum_{n=0}^{N-1} x(n)e^{-i2\pi k\frac n N}$..............................................the Discrete Fourier transform,

\vspace{0.4cm}
$x(n) = \sum_{k=0}^{N-1} X(k)e^{i2\pi k \frac n N}$.........................................................................and its inverse.\\
\end{center}

\newpage
\section*{Act I}
\begin{center}
\emph{in which we see the characters in their formative years}  
\end{center}

\section{Fourier}

In Chapter III of his \emph{Th\'{e}orie Analytique de la Chaleur} \cite{Fourier22,Fourier78} 
from 1822, J. B. Fourier introduces the definition of the series
that bears his name. He develops its formulation as two series of cosines and sines, which
are shown to be independent of each other, one representing even functions and the other
odd functions, respectively.  The discussion culminates in \S 234, where we see 
the enunciation of the equation\endnote{For these quoted equations, an attempt was made to preserve the
original typography and notation from primary sources, for reasons of historical record, hence the variations seen
throughout the paper; on the other hand, 
equations derived by the author are all shown in a modern typographical and notational form.} \\

\begin{equation}\label{eq:fourier_series}
\begin{aligned}
rf(x) =& \frac 1 2 \int_{-r}^{+r} f(x) dx \\ 
&+ \cos \frac {\pi x} r \int f(x)\cos \frac {\pi x} r dx + \cos \frac {2\pi x}r\int f(x) \cos \frac {2\pi x} r dx + etc.\\ &
+ \sin \frac {\pi x} r \int f(x)\sin \frac {\pi x} r dx + \sin \frac {2\pi x} r \int f(x) \sin \frac {2\pi x} r dx + etc.
\end{aligned}
\end{equation}\\

\noindent marked as Eq.(P) in his manuscript. This is followed by the two complementary
half-series, \\

\begin{equation}\label{eq:fourier_series_cos}
\begin{aligned}
\frac 1 2 rf(x) =& \frac 1 2 \int_{0}^{+r} f(x) dx \\ 
&+ \cos \frac {\pi x} r \int f(x)\cos \frac {\pi x} r dx + \cos \frac {2\pi x}r\int f(x) \cos \frac {2\pi x} r dx + etc.\\&
\end{aligned}
\end{equation}\\

and \\

\begin{equation}\label{eq:fourier_series_sine}
\begin{aligned}
\frac 1 2 rf(x) =& \sin \frac {\pi x} r \int_{0}^{+r}  f(x)\sin \frac {\pi x} r dx \\ 
&+ \sin \frac {2\pi x} r \int f(x) \sin \frac {2\pi x} r dx + etc.,\\ 
\end{aligned}
\end{equation}\\

\noindent marked as (M) and (N), respectively. Fourier also shows
that the limits of the integration can be taken from $x = 0$ to
$x=2r$, where the whole interval represented by an arbitrary quantity
$X$, by replacing $r$ in the denominator of the arguments to
each cosine and sine term in Eq.(\ref{eq:fourier_series}).

Next in \S 235, Fourier makes the statement that is at the foundation
of his theory, \\

\begin{quote}
Il r\'{e}sulte de tout ce qui a \'{e}t\'{e} d\'{e}monstr\'{e}  dans cette
section, concernant le d\'{e}veloppement des fonctions en s\'{e}ries
trigonom\'{e}triques, que si l'on propose une fonction $f(x)$, dont la
valeur est repr\'{e}sent\'{e}e das un intervalle d\'{e}termin\'{e}, depuis
$x=0$ jusqu'\`{a} $x=X$, par l'ordonn\'{e}e d'une ligne courbe
trac\'{e}e arbitrairement on pourra toujours d\'{e}velopper cette
fonction en une s\'{e}rie qui ne contiendra que les sinus, ou
les cosinus de multiples impairs. On emploiera, pour conna\^{i}tre
les termes des ces s\'{e}ries, les \'{e}quations (M), (N), (P)\endnote{\emph{It 
follows from this that was demonstrated in this section, concerning 
the development of functions in trigonometric series, that if we 
propose a function $f(x)$, whose value is represented in one 
defined interval, from $x=0$ through $x=X$, by the ordinate of 
a curve traced arbitrarily, we shall always be able to develop this
function in one series that contains only sines or cosines of 
multiple arguments. We shall employ, to find the terms of these
series, the equations (M), (N), (O).} [Author's Translation]}. \cite[p.258]{Fourier22} \\
\end{quote}

Effectively, this somewhat bold statement tells us that we can have
any arbitrary $f(x)$ in an arbitrary interval, and replace it by a
trigonometric series such as the one in Eq.(\ref{eq:fourier_series}).
In addition to this, Fourier also remarks that\\

\begin{itemize}
\item The series shown are always convergent for any function $f$ of a real variable $x$.
\item The sum of two arbitrary functions $f(x)$ and $\phi(x)$ corresponds
to a series where each coefficient is a product of the corresponding
coefficients in the series of these two functions.
\item In Eq.(\ref{eq:fourier_series}), if  $r$ becomes infinitely large,
each term becomes an infinitely small integrand, and the series
is transformed into a definite integral.\\
\end{itemize}

Another important point is also made, that Eq.(\ref{eq:fourier_series})
can be represented by \\

\begin{equation}\label{eq:fourier_series2}
F(x) = \frac 1 \pi \int_{-\pi}^{\pi} F(\alpha)d\alpha\left\{ \frac 1 2 + \sum_{i=0}^{\infty} \cos i(x - \alpha)\right\},
\end{equation}\\

\noindent where \\

\begin{equation}\label{eq:fourier_kernel}
\frac 1 2 +\sum \cos i(x - \alpha)
\end{equation}\\

\noindent is a function of $x$ and $\alpha$, which if multiplied by an arbitrary
function $F(\alpha)$ and integrated from $-\pi$ to $\pi$, it becomes $\pi F(x)$.
That particular function that maps $\alpha$ to $x$ is central to Fourier's
theorem, as we will see later. For now, we should note that this was taken
by P.G. Lejeune-Dirichlet  some years later in 1829 and given the form \cite{Dirichlet}\\

\begin{equation}\label{eq:dirichlet_kernel}
\frac 1 2 + \sum \cos i(x - \alpha) = \frac {\sin\left(i + \frac 1 2 \right)(\alpha - x)} {2 \sin \frac 1 2 (\alpha - x)},
\end{equation}\\

\noindent which later became known as the \emph{Dirichlet kernel}. We shall also revisit this later
on in this paper.

\section{Ohm and Helmholtz}\label{Ohm}

The ideas developed by Fourier, embodied in Eq.(\ref{eq:fourier_series}), enter Music Theory through 
the work of H. Helmholtz as exposed in his \emph{On the Sensations of Tone} \cite{Helmholtz}, and founded on principles established by 
G. Ohm. In 1843, Ohm defends the notion that complex tones can be described by a trigonometric series,
which, although already established earlier by D. Bernoulli, M. Poisson, and others, had been perceived by
him to be under attack\endnote{Ohm states that [t]\emph{he aforementioned observations of expert men, based on experience, concerning the true element of tone, seem to overturn everything previously established in this regard, without any reliable alternative being presented. It struck me as if they were calling for a new definition of tone; however, mindful of the old rule that no other causes should be assumed to explain a natural phenomenon than those that are necessary and sufficient, and given my personal inclination to abandon what I had previously acquired knowledge simply because of the allure of a newly emerging concept, I attempted to determine whether the definition of tone, as it has been passed down to us from our ancestors, contains everything necessary and sufficient for a complete explanation of these new facts. As a result of this test, however, the traditional definition of tone was restored to its full rightful place in a way that seems to me worthy of publication, as I will now demonstrate} \cite[pp.517-8]{Ohm}. He then proceeds to define a tone using the Fourier series. }. 
In order to do so, he brings out the Fourier series, now in the form we are accustomed to see \cite[p.519]{Ohm}, \\

\begin{equation}\label{eq:fourier_series3}
\begin{aligned}
F(t) = A_0 + &A_1 \cos\pi\frac t l+A_2 \cos 2\pi\frac t l+ A_3 \cos 3\pi\frac t l + ...\\
&B_1 \sin\pi\frac t l+B_2 \sin 2\pi\frac t l+ B_3 \sin 3\pi\frac t l + ...,
\end{aligned}
\end{equation}\\

\noindent with coefficients $A_n$ and $B_n$ now defined separately \\

\begin{equation}\label{eq:fourier_coefs}
\begin{aligned}
A_0 &= \frac 1 {2l} \int F(t)dt \\
A_1 &= \frac 1 {l} \int F(t)\cos\pi \frac t l dt\quad  B_1 = \frac 1 {l} \int F(t)\sin\pi \frac t l dt \\
A_2 &= \frac 1 {l} \int F(t)\cos2\pi \frac t l dt\quad  B_2 = \frac 1 {l} \int F(t)\sin2\pi \frac t l dt \\
A_3 &= \frac 1 {l} \int F(t)\cos3\pi \frac t l dt\quad  B_3 = \frac 1 {l} \int F(t)\sin3\pi \frac t dt l \\
& u.s.w  \quad\quad\quad\quad\quad\quad\quad\quad u.s.w, 
\end{aligned}
\end{equation} \\

\noindent and integrated from $t=-l$ to $t=l$.

Ohm states that the ear performs the analysis of complex tones into simple sinusoidal partials,
and that the mathematical form is given by Eqs.~(\ref{eq:fourier_series3}) and (\ref{eq:fourier_coefs}).
This is the basis on which Helmholtz establishes his theory. It is interesting to note that he appears first 
to cast some doubt on the validity of the mathematics involved, \\

\begin{quote}
The theorem of Fourier here adduced shews first that it is mathematically possible to 
consider a musical tone as a sum of simple tones, in the meaning we have attached to the
words, and mathematicians have found it convenient to base their acoustic investigations
on this mode of analysing vibrations. But it by no means follows that we are obliged to consider
the matter in this way. We have rather to enquire, do these partial constituents of a musical
tone such as the mathematical theory distinguishes and the ear perceives, really exist in
the mass of air external to the ear? Is this means of analysing forms of vibration which
Fourier's theorem prescribes and renders possible, not merely a mathematical fiction,
permissible for facilitating calculation, but not necessarily having any corresponding actual 
meaning in things themselves? What makes us hit upon pendular vibrations, and none other,
as the simplest elements of all motions producing sound? \cite[p.36]{Helmholtz}\\
\end{quote} 

However, he follows on to indicate that Fourier's ideas have a probable meaning in nature 
because of the fact that the ear \emph{really effects the same analysis} \cite[p.36]{Helmholtz}, 
which was what Ohm indicated in his paper\endnote{This is now accepted to be true; however
we also known that pitch perception, which was at the centre of his concerns, is complex and 
also involves time-domain effects, beyond Ohm's original theory.}. 
Helmholtz also explores the idea that simple tones (or partials) have
an existence \emph{outside of the ear} through the demonstration of physical effects such
as resonance and sympathetic vibrations. He produces what he claims as experimental proof of Ohm's
partial-tone law. Eventually, he places Fourier's theorem as the 
basis of his theory of harmony, \\

\begin{quote}
Ultimately, then, the reason of the rational numerical relations of Pythagoras is to be
found in the theorem of Fourier, and in one sense this theorem may be considered as
the prime source of the theory of harmony\endnote{It is important to note that Helmholtz
did not admit anything but steady tones with a harmonic spectrum in his definition
of a \emph{musical tone}, which simplifies the analysis somewhat. His music theory
was \emph{only} concerned with these and nothing else.}. \cite[p.36]{Helmholtz}\\
\end{quote}

We should also note that, implicit in Ohm's and Helmholtz's use of the Fourier series,
is the assumption the resulting function of time $F(t)$ in Eq.(\ref{eq:fourier_series3}) is periodic, 
repeating every $2l$ seconds. This is not actually something we find explicitly in Fourier's theory.
However, this was shown later in 1874 by J. O'Kinealy to be the case \cite{Russell}.
From Taylor's theorem, a function of two variables may be expanded as\\

\begin{equation}
f(x + \lambda) = e^{\lambda \frac \delta {\delta x}} f(x).
\end{equation}\\

\noindent Therefore, for a periodic function $f(x + \lambda) = f(x)$, we have\\

\begin{equation}
\left(e^{\lambda \frac \delta {\delta x}}  - 1 \right)f(x)= 0,
\end{equation}\\

\noindent whose auxiliary equation roots are $e^{\lambda p} = 1$, 
with $p = \delta / \delta x$.\\

Since $e^{{i2m\pi}} = \cos(2m\pi) + i\cos(2m\pi) = 1$, we have
$p = i2m\pi/\lambda$, with $m = 0, 1, 2, 3, 4 ... \infty$. The
solution for the differential equation gives us\\

\begin{equation}\label{eq:fourier_series4}
\begin{aligned}
f(x) = A_0 &+ a_1 e^{i2\pi/\lambda}x + a_2 e^{i4\pi/\lambda}x  + ...\\
                    & + b_1 e^{-i2\pi/\lambda}x + b_2 e^{-i4\pi/\lambda}x  + ...,
\end{aligned}
\end{equation}\\

\noindent which is Eq.(\ref{eq:fourier_series3}) expressed using 
complex exponentials. In this form we have 

\begin{equation}\label{eq:fourier_coefs1}
\begin{aligned}
a_n &= \int f(x)e^{-i2n\pi x/\lambda}dx\\
b_n &=  \int f(x)e^{i2n\pi x/\lambda} dx, 
\end{aligned}
\end{equation}\\

\noindent integrated over 0 to $\lambda$, with $i$ now  denoting $\sqrt(-1)$. \\

\section{Fourier}\label{Fourier2}

The integral form of Fourier's series alluded to earlier is probably the most general
statement of his mathematical ideas, and is what is normally referred to as the 
\emph{Fourier theorem}, sometimes also qualified as the double integral bearing his
name. It is developed in Chapter IX, in a similar manner to how Eq.(\ref{eq:fourier_series})
was produced. 

Taking two functions, $F(x)$ and $f(x)$, satisfying $F(x) = F(-x)$ and $f(x) = -f(-x)$,
respectively. The two integrals, \\

\begin{equation}
\frac 1 2 \pi F(x) = \int_0^{\infty} dq \cos qx  \int_0^{\infty}  dx F(x) \cos qx
\end{equation}\\

and \\

\begin{equation}
\frac 1 2 \pi F(x) = \int_0^{\infty} dq \sin qx  \int_0^{\infty}  dx f(x) \sin qx
\end{equation}\\

\noindent are the half-transforms that can be used to arrive at $\phi(x) = F(x) + f(x)$. These
are exactly the result of extending the range of summation in Eqs.~(\ref{eq:fourier_series_cos}) 
and~(\ref{eq:fourier_series_sine}) to infinity as indicated by Fourier earlier in Chapter III.

This is the basis of Fourier's theorem, first given as

\begin{equation}\label{eq:fourier_integral}
\begin{aligned}
\pi \phi(x) = \int_0^\infty &dq \sin qx \int_{-\infty}^{\infty}d\alpha \phi(\alpha)\sin q\alpha \\
+ &\int_0^\infty dq \cos qx \int_{-\infty}^{\infty}d\alpha \phi(\alpha)\cos q\alpha,
\end{aligned}
\end{equation}\\

\noindent which as can be seen introduces a separate variable $\alpha$ in the right-hand-side $\phi()$.
We can view this as the \emph{input} to the expression. The reason for this is that his aim is to develop
an equation like Eq.(\ref{eq:fourier_kernel}) as an integral sum. This he does in the following, labelled (E) in
\S~361, \\

\begin{equation}\label{eq:fourier_theorem}
\phi(x) = \frac 1 \pi \int_{-\infty}^{+\infty} d\alpha \phi(\alpha) \int_0^{\infty} dq \cos q(x - \alpha),
\end{equation}\\

\noindent to which he adds the remark, \\

\begin{quote}
Ainsi la fonction repr\'{e}sent\'{e}e par l'integrale d\'{e}finie $\int dq \cos q(x - \alpha)$
a cette singuli\`{e}re propri\'{e}t\'{e}, que si on la multiplie par une
fonction quelconque, $\phi(\alpha)$ et par $d\alpha$, et si l'on int\`{e}gre par
rapport \`{a} $\alpha$ entre des limites infinies, le r\'{e}sultat est \'{e}gal \`{a}
$\pi\phi(x)$; en sorte que l'effet de l'int\'{e}gration est de changer $\alpha$ en
$x$ et de multiplier par le nombre $\pi$.\endnote{\emph{This way the function represented by the definite
integral $\int dq \cos q(x - \alpha)$ has a singular property, that if we multiply it by
any function whatsoever, $\phi(\alpha)$ abd by $d\alpha$, and if we integrate it with
regards to $\alpha$ between infinite limits, the result is equal to $\pi\phi(x)$; it follows that
the effect of the integration is to change $\alpha$ into $x$ and multiply it by $\pi$.} [A.T.]} \cite[p.449]{Fourier22}\\
\end{quote} 

The significance of this statement in various applications became apparent as Fourier's
work was followed by others. An example of this was given by A. De Morgan, where
Eq.(\ref{eq:fourier_theorem}) is used to describe any function \emph{varying continuously or
discontinuously, in any manner whatever, from $x=-\infty$ to $x=\infty$}. To show this, he
starts by putting Fourier's theorem in the form \cite[p.628]{DeMorgan}

\begin{equation}\label{eq:fourier_theorem_demorgan}
\phi(x) = \frac 1 \pi \int_{0}^{\infty} d\omega \int_{-\omega}^{+\omega} \cos \omega(\nu - x)\phi(\nu) d\nu.
\end{equation}\\

Following this, he sets up $\phi(\nu)$ as made of three segments, $\phi(\nu) = 0$, from $\nu = -\infty$ to
$\nu = a$; $\phi(\nu)= \psi(\nu)$, from $\nu = a$ to $\nu = b$; and $\phi(\nu) = 0$, from $\nu = b$ to
$\nu = \infty$. Now he replaces $-\omega$ and $+\omega$ by $a$ and $b$. Using the constant
function $\psi(\nu) = 1$, we have

\begin{equation}\label{eq:fourier_theorem_demorgan2}
\phi(x) = \frac 1 \pi \int_{0}^{\infty} d\omega \int_{a}^{b} \cos \omega(\nu - x) d\nu.
\end{equation}\\

With this, he arrives at

\begin{equation}\label{eq:demorgan_rect}
\frac 1 \pi \int_0^{\infty} \frac {\sin(b - x)\omega}\omega d\omega - \frac 1 \pi \int_0^{\infty} 
\frac {\sin(a - x)\omega}\omega d\omega = \begin{cases}
0, & -\infty < x < a  \\
\frac 1 2, &x = a \\
1, &a < x < b \\
\frac 1 2, &x = b \\
0, &x = b < x < \infty, \\
\end{cases}
\end{equation}\\

\noindent which describes what is now known as a rectangular function. We will later revisit this in
another guise.

De Morgan follows this by showing that with Fourier's theorem in the form

\begin{equation}\label{eq:fourier_theorem_demorgan3}
\frac 1 \pi \int_{0}^{\infty} d\omega \int_{a}^{b} \cos \omega(\nu - x)\phi(\nu) d\nu,
\end{equation}\\

\noindent which he labelled $F_a^b \phi(x)$, e.g. $F_a^b 1$ for Eq.(\ref{eq:demorgan_rect}), we may describe
any function in segments, $a$ to $b$, $b$ to $c$, $c$ to $d$, etc, as the sum\\

\begin{equation}\label{eq:fourier_theorem_demorgan4}
f(x) = F_a^b \phi(x) + F_b^c \psi(x) + F_c^d \chi(x) + etc .
\end{equation}\\

\noindent In this case, we have $f(a) = \frac 1 2 \phi(a)$,  $f(b) = \frac 1 2 [\phi(b) + \psi(b)]$, 
$f(c) = \frac 1 2 [\psi(c) + \chi(c)]$, etc.

De Morgan also provides another example where we have a ramp $f(x) = x$ from 0 to 1, zero elsewhere,
and notes that the result is the limit of\endnote{A. Freeman in his translation of Fourier's book notes
that this equation has a proof attributed to Poisson \cite{Poisson}, which many authors regard as
the definite proof of Fourier's double integral \cite[p.351]{Fourier78}.}\\

\begin{equation}\label{eq:fourier_theorem_demorgan5}
\frac 1 \pi \int_{0}^{\infty} e^{-k\omega} d\omega \int_{a}^{b} \cos \omega(\nu - x) . \nu d\nu,
\end{equation}\\

\noindent as $k \rightarrow 0$, except for the discontinuity point at $x = 1$, where it is $1/2$ \cite[p.628]{DeMorgan}. 

We should remark that this application of Fourier's theory is quite distinct from that of Ohm and Helmholtz. Here, instead
of exploring periodic functions, the focus is on the finding analytical expressions for discontinuous cases.
To bridge these two perspectives, we can say that this work gives us the means of understanding, for example, a
waveform that is turned on and off, a single cycle that is not repeated, etc. This has implications for, amongst other things,
our understanding of time and frequency in the context of the Fourier transform, and the possibility of slicing functions 
at will to manipulate them. We shall return to this later in the paper.

\section{Dirac}

Further tools for handling these emerging questions about infinity are provided by P.A.M. Dirac's \emph{delta}.
A notation for this was first given in his 1927 paper\endnote{It is interesting to note how in the 20th century
the paths of Quantum Physics and the theory of musical signals intertwine. However, one should have guessed
that, as both can be seen as branches of Mathematical Physics.} 
\emph{The Physical Interpretation of the Quantum Dynamics} \cite{Dirac1927}, where he introduces 
the symbol $\delta(x)$, \\

\begin{equation}\label{eq:delta1}
\delta(x) = 0 \quad \textrm{when} \quad x \neq 0
\end{equation} \\

\noindent and\endnote{In Dirac's original paper this is given as $\int x\delta(x) = 1$, we use
here instead the definition provided in his later work 
\emph{The Principles of Quantum Physics} \cite{Dirac1947}.}\\
 
\begin{equation}\label{eq:delta2}
\int_{-\infty}^{\infty} \delta(x) dx = 1. 
\end{equation} \\

Dirac states at the outset that $\delta$ is not a proper function of $x$, and we shall continue to observe
this by omitting the word \emph{function} in relation to it. He also remarks that it may be seen
\emph{as a limit of a certain sequence of functions}. The main idea of this representation is to give us
a form to treat a mathematical object that is zero everywhere except at an infinitely small area around 0.

There are a number of important properties that are relevant in the context of this paper. The first one
of these is

\begin{equation}\label{eq:delta3}
\int_{-\infty}^{\infty} f(x)\delta(x) dx = f(0),
\end{equation} \\

\noindent for any continuous $f(x)$. From the earlier definitions, it seems clear that the 
product by $\delta(x)$ and integration effectively select the value of the function for $x=0$.

We can now deduce 

\begin{equation}\label{eq:delta4}
\int_{-\infty}^{\infty} f(x)\delta(x-a) dx = f(a)
\end{equation} \\
 
\noindent with $a$ standing for any real number. Dirac makes the following remark in
relation to this equation:\\

\begin{quote}
the process of multiplying a function of $x$ by $\delta(x-a)$ and integrating over all $x$ is equivalent
to substituting $a$ for $x$. \cite[p.59]{Dirac1947}\\
\end{quote}

This brings us back to Fourier's theorem, where a similar statement was made about the definite integral
in Eq.(\ref{eq:fourier_theorem}). Indeed, we can now spot that \\

\begin{equation}\label{eq:fourier_delta}
\int_0^{\infty} dq \cos q(x - \alpha) = \pi \delta(x - \alpha),
\end{equation} \\

A proof is fairly straightforward:

\begin{enumerate}
\item For the case where $x \neq \alpha$, the left hand-side of Eq.(\ref{eq:fourier_delta}) vanishes\endnote{This 
follows Dirac's approach where \emph{an integral whose value to the upper limit $q'$ is of the form
$\cos aq'$ or $\sin aq'$, with $a$ a real number not zero, is counted as zero when $q'$ tends to infinity,
i.e. we take the mean value of the oscillations} \cite[p.95]{Dirac1947}.}, so
we have $\delta(x) = 0$ for $x \neq 0$.
\item With $x = \alpha$, we can revert to Eq.(\ref{eq:fourier_theorem}), where we have

\begin{equation}\label{eq:fourier_delta_proof}
\frac 1 \pi \int_{-\infty}^{+\infty} \phi(x) \left\{\int_0^{\infty} \cos q(0) dq  \right\} dx = \phi(0),
\end{equation}\\

\noindent which can be compared to Eq.(\ref{eq:delta3}) to prove $\delta(x)$ for $x = 0$ and
$\pi^{-1} \int \cos q(x) dq = \delta(x)$.\\ 
\end{enumerate} 

With this, the connection between Fourier's theory and the infinities described by the $\delta$
are made evident. Dirac sees the $\delta$ as an \emph{improper function} without a well-defined value. 
However, when appearing as a factor in an integrand, it leads to the integral
having a well-defined value. Therefore its appearance should not be regarded as a lack of
rigour in this case, but as a convenient notation. 

Two final aspects of the $\delta$ are worth noting as they have relevance to the matter at
hand. Firstly, the $\delta$ can also be defined as the derivative of the step function,
originally introduced by O. Heaviside \cite[p.510]{Heaviside}, and quoted by Dirac as\endnote{We should notice
how the difficult question of $x=0$ is side-stepped here, pun excused.}

\begin{equation}\label{eq:step}
\epsilon(x) = 
\begin{cases}
0,  &\quad x < 0 \\
1, &\quad x > 0.
\end{cases}
\end{equation} \\

\noindent and as proof $\epsilon'(x)$ can be replaced for $\delta(x)$ in Eq.(\ref{eq:delta3}) and
it can be solved to get as a result $f(0)$\endnote{We can also note the origin of Dirac's delta
given by Heaviside as $p^1 = \frac 1 \pi \int_0^{\infty} \cos \upsilon t d\upsilon$, which he
notes as being \emph{the basis of Fourier's theorem}, but more convenient to use \cite[p.510]{Heaviside}. 
His is yet another proof of Eq.(\ref{eq:fourier_delta}), ahead of time.}. 
This demonstrates that a delta is involved whenever
we try to differentiate a discontinuity.

The second point is that $\delta(x)$ can be differentiated $n$ times, resulting in $\delta'(x)$,
$\delta''(x)$, and so on. Associated with this we have the fact that while, trivially, 
$\delta(x) = \delta(-x)$, therefore even, its derivative is odd, $\delta'(x) = -\delta'(-x)$. Following
this, we can also deduce that the second-order derivative is even, the third is odd, and so on.
Any derivative of order $n$, $\delta^{(n)}(x)$ can replace it in Eq.(\ref{eq:delta4}) producing as a result the
respective derivative of $f(a)$, $f^{(n)}(a)$.

\newpage

\section*{Act II}
\begin{center}
\emph{in which the characters operate in time and frequency}  
\end{center}

\section{Transforms}

We have now effectively all the tools to explore the Fourier transforms, giving it a physical interpretation
that was mostly missing so far. From Eq.(\ref{eq:fourier_integral}), we can develop a modern 
musically-meaningful form of his eponymous Transform. We will proceed to do this by \\

\begin{itemize}
\item splitting the single equation into two,
\item replacing the real-valued sinusoids by complex exponentials; and
\item employing functions of time $t$ in seconds and frequency $f$ in Hz. \\
\end{itemize}

This gives the pair \\

\begin{equation}\label{eq:transform}
X(f) = \int_{-\infty}^{\infty} x(t)e^{-i2\pi f t } dt,
\end{equation} \\

\noindent and\\

\begin{equation}\label{eq:inverse_transform}
x(t) = \int_{-\infty}^{\infty} X(f)e^{i2\pi ft} df.
\end{equation} \\

In this way, the transforms are set in a very compact form, and it is possible to observe the
relationship between time on side and frequency on the other. We can use the term \emph{spectrum}
as introduced by D. Gabor \cite{Gabor} to denote the frequency domain, and \emph{waveform} for the time domain.
It is also possible to develop an intuition about the operation of the transforms. Firstly, we should interpret 
the complex exponential for what it is in this context, a sinusoid with arbitrary phase and frequency.
Secondly, we can observe that, starting with a waveform defined by $x(t)$, we use a product of it
with a sinusoid of frequency $f$, and by integrating over all $x$ we get a complex number that
tells us the amplitude and phase of the waveform at that frequency. We may even go as far as
using a radio receiver metaphor, the frequency $f$ is what we set in our dial, and if there is
anything there in the waveform, we will get a result.

From an inverse perspective, once we have $X(f)$ defining amplitudes and phases in the
spectrum, we can recompose the waveform by multiplying these with complex sinusoids
at all frequencies $f$, then effecting an integral sum of all these partial waveforms. This
can be described as a form of \emph{additive synthesis}. Additionally, having the 
spectrum $X(f)$, allows us to pick individual partials if we wish, instead of applying the
full summation.

\section{Zero Frequency and Zero Time}

With that in mind, we can now exact the mathematical form in which this can be represented.
Starting with the simplest of the waveforms, one which is just a constant, $x(t) = 1$ for all
$x$, we have \cite[p.95]{Dirac1947}\\

\begin{equation}\label{eq:transform_dc}
X(f) = \int_{-\infty}^{\infty} e^{-i2\pi f t } dt = \delta(f).
\end{equation} \\

\noindent This result can be verified in various ways, intuitively first by noting that both imaginary and real
parts of this equation vanish for $x \neq 0$, and at $X(0)$ the limit of the sum is 1. Secondly, we can replace
it in Eq.(\ref{eq:delta3}), and solve it to get exactly $X(0)$, and finally, we can use Fourier's theorem and
Eq.(\ref{eq:fourier_delta_proof})\endnote{Dirac gave Eq.(\ref{eq:transform_dc}) as 
$\int_{-\infty}^{\infty} e^{iax} dx = 2\pi\delta(a)$, i.e. in the form of the inverse transform, and proved
it using yet another route, 
$\int_{-\infty}^{\infty} f(a) d(a) \int_{-g}^{g} e^{iax} dx = \int_{-\infty}^{\infty}f(a)da\, 2a^{-1} \sin ag = 2\pi f(0)$,
for $g \rightarrow \infty$ \cite[p.95]{Dirac1947}.}.

The physical interpretation is as follows, since for $f = 0$, we have $\cos(2\pi f t) = 1$ (and $\sin(2\pi f t) = 0$),
this is a signal with energy concentrated in one single frequency, 0Hz. We commonly call this, with
reference to electrical signals, a \emph{direct current} (DC) waveform. We should note the exact role of the delta in this 
context, as a representation of the fact that waveform amplitude is fully concentrated at a single point
in the frequency axis.

Conversely, let's consider a signal that is concentrated at one point in time, and nowhere else, an
infinitesimally small \emph{pulse}. This can be represented by $x(t) = \delta(t)$,
and we should very well expect a similar result by employing the inverse transform,\\

\begin{equation}\label{eq:inverse_transform}
x(t) = \int_{-\infty}^{\infty} e^{i2\pi ft} df = \delta(t).
\end{equation} \\

\noindent That is, if the energy of a waveform is concentrated at a single time point, then it is
unity everywhere in the frequency domain, $X(f) = 1$. These two results should be regarded as
the foundation for handling the duality of time and frequency. Another way of putting this is 
to say that if a musical signal is defined precisely in time, it cannot be determined in frequency;
conversely if it is exactly defined in frequency, it cannot be determined in time. This result
is of course a fundamental fact in Quantum Physics in the form given by Heisenberg's 
\emph{principle of uncertainty} \cite[p.98]{Dirac1947}, as well as in the theory of musical 
signals as interpreted by Gabor \cite{Gabor}\endnote{This principle can be stated 
by the identity $\Delta t \Delta f \simeq 1$, in which $\Delta f$ stands for the frequency 
bandwidth and $\Delta t$ for the time period of the analysis.}.

\section{Sinusoids}

Now we can extend these results sideways, so to speak, taking advantage of Eq.(\ref{eq:delta4}).
Let's consider a real-valued sinusoid with amplitude $A$, frequency $F$, and phase $\theta$, which
can be described by\\

\begin{equation}\label{eq:sinusoid}
x(t) = A\cos(2\pi F t + \theta) = A[a\cos(2\pi Ft) - b\sin(2\pi Ft)]
\end{equation} \\

Applying the transform, and using Eqs.~(\ref{eq:transform_dc}) and~(\ref{eq:delta4}), 
we get \\

\begin{equation}\label{eq:sinusoid_transform}
\begin{aligned}
X(f) &= \int_{-\infty}^{\infty} A[a\cos(2\pi F t) - b\sin(2\pi F t)] e^{-i2\pi ft} dt \\ 
&= \frac A 2 \left[a\{\delta(f - F) + \delta(f + F)\} + ib\{\delta(f - F) - \delta(f + F)\}\right],
\end{aligned}
\end{equation} \\

\noindent that is, energy concentrated at two points in the frequency axis, at $f$ and
$-f$, each with half amplitude\endnote{This result shows how the Dirac delta is useful
as a notation to denote something that would be otherwise awkwardly described using
Fourier's theorem for each term.}. Notice that the result is generally complex, and the values of $a$ and $b$ 
determine the phase of the sinusoid as per the usual formula for converting 
rectangular to polar forms of complex numbers. The spectra of real-valued
waveforms is always split into negative and frequency sides, which are symmetric
around 0 Hz for their real part, and anti-symmetric for their imaginary part. This
type of symmetry is called \emph{Hermitian} \cite[p.68]{Dirac1947}. From this result, we can deduce that
if the waveform is made up of a mixture of various sinusoids, its spectrum will
be a sum of deltas at each absolute frequency, with half of their amplitudes on the 
positive side and half on the negative. We should expect it to be zero elsewhere\endnote{This of course assumes 
a waveform that exists ad infinitum. As noted by D. Gabor, \emph{the Fourier-integral method considers
phenomena in an infinite interval, sub specie aeternitatis, and
this is very far from our everyday point of view} \cite[p.431]{Gabor}.
}.

The converse case is also interesting, what is the time domain function corresponding to
a spectrum consisting of a single complex exponential, $X(f) = e^{-2\pi \tau  f}$? We can
proceed along similar lines to obtain \\

\begin{equation}\label{eq:complexe_transform}
x(t) = \int_{-\infty}^{\infty} e^{2\pi (t-\tau)f} df = \delta(t-\tau),
\end{equation} \\

\noindent which can be given the physical interpretation of a time delay. This
makes the pulse appear at time $t=T$ rather than at $t=0$. In fact, a relevant notion
that arises from this, intuitively, is that if we take the product of an arbitrary spectrum
and a complex exponential such as this, we will be applying a time delay of sorts
to a waveform.

\section{Series}\label{Series}

What was said about the energy concentration in a waveform that consists solely 
of sinusoids can now be extended to give a generic formula for the Fourier series.
If we have a waveform composed of sinusoids at exact multiples of a frequency $F$,
then a discrete sum can be used to reconstitute it from its spectrum $X(f)$,

\begin{equation}\label{eq:fourier_series_generic}
x(t) = \sum_{k=-\infty}^{\infty} X(kF)e^{i2\pi kFt},
\end{equation} \\

\noindent since we know this is composed purely of a sum of $\delta(f \pm kF)$.
Moreover, because of the earlier results in \S \ref{Ohm}, we should expect
that the waveform is periodic over a period $T = \frac 1 F$ and so the spectrum
only needs to be computed over this interval, for discrete frequencies $nF$,

\begin{equation}\label{eq:transform_discrete}
X_F(n) = \int_{-\frac T 2}^{\frac T 2} x(t)e^{-i2\pi nF t } dt,
\end{equation} \\
 
\noindent which certainly simplifies things. This pair of transform-series 
computes a waveform from a discrete sequence of evenly-spaced 
frequency values.

Completing the set, we have a transform that takes an evenly-spaced 
discrete sequence of waveform values, and computes its continuous-frequency
spectrum,

\begin{equation}\label{eq:discrete_time_transform}
X(f) = \sum_{n=-\infty}^{\infty} x(nT)e^{-i2\pi fnT}
\end{equation} \\

\noindent for which, we should expect the inverse transform to be an integral
over an interval $F = \frac 1 T$,\\

\begin{equation}\label{eq:transform_discrete}
x_T(n) = \int_{-\frac F 2}^{\frac F 2} X(f)e^{i2\pi nf T } dt,
\end{equation} \\

\noindent producing a sequence of time values at evenly-spaced intervals.

These results point to the general principle that the spectrum of a discrete waveform is 
periodic in frequency, with period $F$ corresponding to the reciprocal of the 
time interval $T$ between points in the series. Conversely, the waveform computed from a discrete
spectrum is also periodic, with period $T$, corresponding to the reciprocal of the 
frequency interval $F$ between points in the series. We shall return to this later as it gives 
the foundation for a fully discrete transform pair.

\section{Dirac Comb}

At this stage, we can expand our results by introducing another useful notation as
an aid for more complicated cases. Reverting to the Dirichlet kernel introduced
earlier in Eq.(\ref{eq:dirichlet_kernel}), if we rewrite it using complex exponentials, 
we have\\

\begin{equation}\label{eq:dirac_comb}
x(t) = \sum_{n=-\infty}^{\infty} e^{2\pi nt }  = \sum_{n=-\infty}^{\infty} \delta(t-n) = \Sha(t),
\end{equation} \\

\noindent which will be referred to as the \emph{Dirac comb}, denoted by $\Sha$\endnote{The Cyrillic
alphabet letter \emph{sha}. This appears to be derived from the Hebrew \emph{shin}, which 
is known in some places to mean \emph{live long and prosper}.}. Similarly to the delta, we
should refrain from it calling a function, and our use in this context will be simply as 
an operator in expressions (as demonstrated later). Heuristically, it can be interpreted
as an \emph{impulse train}.

An interesting property of the comb is this, \\

\begin{equation}\label{eq:comb_transform}
\begin{aligned}
X(f) &= \int_{-\infty}^{\infty} \sum_{n=-\infty}^{\infty} \delta(t - n) e^{-2\pi tf} dt = \sum_{n=-\infty}^{\infty} e^{-i2\pi fn} \\
&= \sum_{n=-\infty}^{\infty} \delta\left(f - n\right) = \operatorname{III}(f),
\end{aligned}
\end{equation} \\

\noindent which tells us that the spectrum of a comb waveform is also a comb, 
and conversely.

\section{Convolution}\label{Convolution}

It is useful now to determine what is the frequency domain function corresponding
to the product of two waveforms. Starting with the expression $x(t) = g(t)h(t)$,
and applying the forward transform of Eq.~(\ref{eq:transform}), we obtain \\

\begin{equation}\label{eq:convolution}
\begin{aligned}
X(f) &= \int_{-\infty}^{\infty} x(t) e^{-i2\pi ft} dt = \int_{-\infty}^{\infty} g(t)h(t) e^{-i2\pi ft} dt\\
&= \int_{-\infty}^{\infty} g(t) \int_{-\infty}^{\infty} H(\phi)e^{i2\pi \phi t}d \phi e^{-i2\pi ft} dt \\
&= \int_{-\infty}^{\infty}\int_{-\infty}^{\infty} H(\phi) g(t) e^{-i2\pi(f-\phi)t} dt d\phi 
= \int_{-\infty}^{\infty} H(\phi) G(f-\phi) d \phi,
\end{aligned}
\end{equation} \\

\noindent which introduces an operation called the \emph{convolution} of $H(f)$ and $G(f)$.
Well-known trigonometric identities such $\cos(\omega)\cos(\theta) = \frac 1 2 [\cos(\omega - \theta) +
\cos(\omega + \theta)]$ follow from this result. It should also not surprise us that the product of 
two spectra corresponds to the convolution of two waveforms, which can be demonstrated by 
the same methods applied here.

What is the physical interpretation of these results? We can observe that there are three 
separate operations wrapped in the convolution,\\

\begin{enumerate}
\item translation;  
\item product; and
\item (integral) sum.\\
\end{enumerate}

\noindent This should give us an intuition as to what is taking place here. We \emph{slide}
one function continuously for all $\phi$, taking the products of this and the other function
at $f$, then we sum it all together. We can think of this as an accumulation of \emph{copies}
of one function, scaled by another at every infinitesimal point of the domain. Or simply put,
a \emph{smear} over frequency or over time. This becomes much simpler to picture if
we apply it to discrete sequences, as the integral becomes just a sum.

\section{Discrete Time and Discrete Frequency}

We now have the means to go from continuous to discrete time in a more rigorous way. 
By reasoning that the process of forming a sampled\endnote{The word \emph{sample}
here was introduced by C. Shannon \cite{Shannon} to describe the process of taking values, \emph{samples},
off a continuous-time function representing a signal of some kind.}  discrete sequence of evenly-spaced values
from a continuous function $x(t)$ can be equated to applying a set of deltas at each
point, we can then apply a Dirac comb as an operator to represent this process\endnote{The 
use of the Dirac comb here as if we were taking its product with a function can be
seen as abuse of notation. In fact, what is happening here is that we have a sum of
$\int x(t)\delta(t - nT) dt$ as a result of the operation. Therefore we opt to call this
not a product but the application of the Dirac comb as a \emph{sampling} operator,
which selects the values of $x(t)$ at integer multiples of $T$. The result is a continuous
time function that is zero except (possibly) at these sampling points.},\\

\begin{equation}\label{eq:sampling_time}
x_T(t) = x(t) \frac 1 T\sum_{n=-\infty}^{\infty} \delta(t - nT)= x(t)\Sha_{T}(t),
\end{equation} \\

\noindent which may be simply referred to as \emph{sampling}. The underscore $T$
now refers to the sampling interval.

Intuitively, from \S \ref{Convolution} we should expect the spectrum of this sequence to 
be a convolution of the spectra of $x(t)$ and the Dirac comb. This can be
verified by

\begin{equation}\label{eq:spectrum_sampled}
\begin{aligned}
X_T(f) &= \int_{-\infty}^{\infty} x_T(t) e^{-2\pi ft} dt\\  
&= \int_{-\infty}^{\infty} X\left(f - \phi\right)\Sha_{T^{-1}}(\phi) d \phi,  
\end{aligned}
\end{equation} \\

\noindent where $T^{-1}$ is now known as the sampling frequency. It also confirms
the general principle from \S \ref{Series}, that the spectrum of a discrete time series
is periodic in frequency.

The physical interpretation of this is as follows. A periodically sampled function of
time has a spectrum that repeats at intervals of the sampling frequency $F$, which can
be thought of as images of a base frequency band ranging from $-\frac F 2$ to $\frac F 2$.
This implies that we only need to look at this base band to know what the spectrum
of a sampled waveform is.

Now, using the same principles, we can determine what happens when we sample
a spectrum instead,

\begin{equation}\label{eq:sampling_frequency}
X_F(t) = X(f) \frac 1 F\sum_{n=-\infty}^{\infty} \delta(t - nF)= X(f)\Sha_{F}(f),
\end{equation} \\

\noindent and reconstitute a waveform from it,

\begin{equation}\label{eq:waveform_sampled}
\begin{aligned}
x_F(t) &= \int_{-\infty}^{\infty} X_F(f) e^{2\pi ft} dt\\  
&= \int_{-\infty}^{\infty} x\left(t - \tau\right)\operatorname{III}_{F^{-1}}(\tau) d \tau . 
\end{aligned}
\end{equation} \\

This is effectively the result of the series discussed in 
\S \ref{Series} put in a more generic form. It tells us precisely that the sampled
spectrum produces a periodic waveform with period $F^{-1}$. We should note
that these equations do not imply that the underlying continuous-time 
waveform $x(t)$ has a spectrum that is band-limited to an interval $F$ or that
the spectrum $X(f)$ corresponds to a waveform that is time-limited to an
interval $T$. If they are not, there will be an overlap, in frequency or in
time, between the \emph{repeated copies} of the spectrum or waveform.
This effect is called \emph{aliasing}\endnote{Shannon \cite{Shannon} showed
that if a spectrum has energy only in the base band in an interval $F = \frac 1 T$,
extending from $-\frac F 2$ to $\frac F 2$, then the corresponding continuous-time
waveform can be completely described by its samples. This is a time dual of Fourier's 
theorem. In this case, if a continuous waveform extends over an interval $X = \frac 1 F$ 
and is zero elsewhere, then its continuous-time spectrum can be completely
described by its samples. In both cases, aliasing is avoided.}

\section{Finite Time and Finite Frequency}

We should now turn our attention to defining what goes on when we turn a waveform
on and off. This reverts back to the ideas in \S \ref{Fourier2}, where a focus was placed
on this exact issue. We may equate the turning on/off of a waveform to the application
of a function that is 1 for a given time interval and zero elsewhere, for a which an
expression was derived in Eq.(\ref{eq:demorgan_rect}). With such interval defined
evenly around the origin as $T$, this is commonly known as the rectangular 
function $\Pi_T(t)$. For $T=1$ we then have, \\

\begin{equation}\label{eq:rect}
\begin{aligned}
\Pi(t) = \begin{cases}
0, & -\infty < t < 1/2  \\
\frac 1 2, &t = | 1/2| \\
1, &1/2 < t < 1/2 \\
0, &t = 1/2 < x < \infty \\
\end{cases} &= \frac 1 \pi \int_0^{\infty} \frac {\sin\left(\left(\frac 1 2 - t\right)\omega\right)}\omega d\omega 
- \frac 1 \pi \int_0^{\infty} \frac {\left(\sin\left(t-\frac 1 2\right)\omega\right)}\omega d\omega \\
&= \frac 2 {\pi} \int_{0}^{\infty} \frac {\sin(\omega)\cos(\omega t)} \omega  d\omega =
\frac 1 {\pi} \int_{-\infty}^{\infty} \frac {\sin(\omega)} \omega e^{\omega t} d\omega =\\ 
&= \int_{-\infty}^{\infty} \frac {\sin(\pi f)} {\pi f} e^{2\pi ft} df = \int_{-\infty}^{\infty} \operatorname{sinc}(f)e^{2\pi ft} df,
\end{aligned}
\end{equation} \\

\noindent which tells us that its spectrum is a sinc function. Similarly, it would not be again a surprise
to see that the time domain waveform corresponding to a rectangular spectrum is also a sinc function.

We can now apply this to our problem of switching on/off a waveform\endnote{The full definition for the rectangle function with an arbitrary positive $T$ is $\Pi_T(t) = \Pi\left(\frac t T\right)$.}, \\

\begin{equation}\label{eq:onoff}
x_T(t) = x(t)\Pi_T(t)
\end{equation} \\

\noindent which corresponds to the spectrum\endnote{Likewise, $\operatorname{sinc}_T(\phi) = T \operatorname{sinc}(T\phi) = \frac {\sin(\pi T\phi)} {\pi\phi}$.} 

\begin{equation}\label{eq:onoff_spectrun}
\begin{aligned}
X_T(f) &= \int_{-\infty}^{\infty} x_T(t) e^{-2\pi ft} dt\\  
&= \int_{-\infty}^{\infty} X\left(f - \phi\right)\operatorname{sinc}_T(\phi) d \phi  
\end{aligned}
\end{equation} \\

Conversely, we can also cut the spectrum in a given band using the rectangular
function,

\begin{equation}\label{eq:band}
X_F(t) = X(f)\Pi_F(f)
\end{equation} \\

\noindent resulting in the waveform

\begin{equation}\label{eq:band_waveform}
\begin{aligned}
x_F(t) &= \int_{-\infty}^{\infty} X_F(f) e^{2\pi ft} dt\\  
&= \int_{-\infty}^{\infty} x\left(t - \tau\right)\operatorname{sinc}_F(\tau) d \tau  
\end{aligned}
\end{equation} \\

The physical interpretation of this allows us, amongst other things, to treat the effects of aliasing described earlier. If we
take a continuous waveform and apply the convolution to a sinc function, the resulting spectrum is
the product of a rectangular function and the waveform spectrum. This therefore eliminates all
frequencies outside a given base band. When we sample the resulting waveform we will be sure
that no frequency aliasing effects are due. Conversely, if we take the product of a rectangle with 
width $T$ and a given waveform, the resulting spectrum is the convolution of a sinc function and the spectrum
of the waveform. We can then sample that spectrum at multiples of $F = \frac 1 T$ without any
concern for time aliasing.

To give a complete picture of our current state of affairs, 
we can combine the operations of sampling and cutting waveforms in
one single expression,

\begin{equation}\label{eq:samplecutwave}
x_{TS}(t) = x(t)\Sha_T(t)\Pi_S(t).
\end{equation} \\

\noindent This corresponds to the continuous spectrum

\begin{equation}\label{eq:samplecutspectrum}
\begin{aligned}
X_{TS}(f) &= \int_{-\infty}^{\infty} x_{TS}(t) e^{-2\pi ft} dt\\  
&=  \int_{-\infty}^{\infty} X_T\left(f - \phi\right)\operatorname{sinc}_S(\phi) d \phi,  
\end{aligned}
\end{equation} \\

The sampled, band-limited spectrum defined as

\begin{equation}\label{eq:sampleblspectrum}
X_{FS}(f) = X(f)\Sha_F(f)\Pi_S(f),
\end{equation} \\

\noindent produces the waveform 

\begin{equation}\label{eq:sampleblwave}
\begin{aligned}
x_{FS}(t) &= \int_{-\infty}^{\infty} X_{FS}(f) e^{2\pi ft} df\\  
&=  \int_{-\infty}^{\infty} x_F\left(t - \tau\right)\operatorname{sinc}_S(\tau) d \tau  
\end{aligned}
\end{equation} \\

\section{Discrete Transforms}

If we can cut a slice of a sampled waveform as a sequence of $N$ values, then using the
ideas developed earlier, we can obtain a spectrum which consists of $N$ frequency
values, each one a complex number. Here is how we put it,

\begin{equation}\label{eq:discrete_transform}
X(k) = \sum_{n=0}^{N-1} x(n)e^{-i2\pi k\frac n N},
\end{equation} \\

\noindent for $k \in \mathbb{Z}$. This completely describes the discrete 
spectrum, which is periodic, with period $N$. The inverse transform, starting with a 
discrete spectrum $X(k)$ with $N$ points, can be put as

\begin{equation}\label{eq:inverse_discrete_transform}
x(n) = \sum_{k=0}^{N-1} X(k)e^{i2\pi k\frac n N}
\end{equation} \\

\noindent for $n \in \mathbb{Z}$. The two transforms can be used transparently,
e.g. forwards and inverse, like the continuous versions, to recover the original sequence,
but a scaling factor $1/N$ needs to be applied once in either direction\endnote{In fact, we
can express these transforms as $\mathbf{X} = \frac 1 N \mathbf{x}\mathbf{F}$ and 
$\mathbf{x} = \mathbf{X}\mathbf{F}^{-1}$, where $\mathbf{x}$ and $\mathbf{X}$ are $N$-sized column and
row vectors, respectively, and $\mathbf{F}$ is an $N$ by $N$ matrix containing the complex sinusoids,
$e^{-2\pi kn/N}$. This form actually allows us to consider factorisations that may help us compute
the transforms more efficiently, using fast algorithms. The most well-known of these is the
fast Fourier transform. Its history of development appears to stretch to C.F. Gauss who 
developed an early form of it as a means of computing the coefficients of a finite 
Fourier series. This was done independently of Fourier, and most likely contemporaneously, 
but went unpublished in Gauss' lifetime \cite{Heideman}.}.

In order for us to link $k$ to frequencies $f$ in Hz, we first set $t = nT$,
where $T$ is the sampling period, so the time series is set at
$0, T, 2T, ..., (N-1)T$ seconds. Thus $F = \frac 1 T$
is our sampling frequency. Then the frequency points $k$ are
set at $0, F/N, 2F/N, ..., (N-1)F/N$ Hz. However, due to the
periodicity in frequency, the points from $N/2$ to $N-1$ also refer
to the frequencies $-F/2$ to $-F/N$, the negative side of 
the base band spectrum. It is more convenient to read
the second half of the discrete transform spectrum as referring to
negative frequencies, rather than their positive images.

The discrete transform may have two interpretations with regards to the underlying
continuous-time functions, both equally valid. From the perspective of the forward
transform, the first one is that we are looking at a waveform that is zero for all $t$, except for
a time window of $NT$ seconds. In this case the underlying continuous-frequency 
spectrum is the convolution of a sinc function and the sampled spectrum. This latter function 
can be represented in continuous time by a sum of scaled deltas at each
sampling point $\frac {kF} {N}$. This is what Eq.(\ref{eq:samplecutspectrum}) 
gives as a result, and is equivalent to the methods explored in \S \ref{Fourier2}.

The second interpretation is that the waveform is a single cycle of a periodic
function of time, which repeats from $t=-\infty$ to $t=\infty$, eternally. 
In this case the underlying continuous-time spectrum is zero everywhere, 
except at the sampling points, where it is effectively the same 
as the sampled spectrum. It is a sum of scaled deltas spaced evenly in frequency. 
While both are equally good representations, the first one, 
in conjunction with the principles from  \S \ref{Fourier2}, allows us to employ the 
discrete transforms to successive blocks of a waveform successfully to obtain 
its spectrum. The second one allows us to create single-cycle waveforms in
oscillators for wavetable sound synthesis.

Similar points can be made to the underlying continuous waveform of a 
sampled spectrum, if we take the spectrum as periodic, with period $F$, 
stretching from $f=-\infty$ to $f=\infty$, then the underlying time-domain waveform 
is exactly the same as a a sequence of scaled deltas at the sampling points. That is, infinitely small 
areas with zeros in between. On the other hand, if we take it as fully band-limited
to $-\frac F 2$ to $\frac F 2$, with $X(f) = 0$ elsewhere, 
then the reconstruction of the continuous-time signal from samples is a convolution 
of sinc functions by these scaled deltas, that is, a smooth waveform. 

In both time and frequency-domain cases, whenever we are taking a slice of the function within
a given interval, we are multiplying it by a rectangular function. The result, as we have seen, is
a convolution of the corresponding waveform or spectrum with a sinc function. Without this, we are
limited to a single interpretation. This tells us that the continuous-time functions in both domains 
are a sum of translated, scaled deltas, periodic, and stretching infinitely.

\newpage

\section*{Epilogue}

\begin{center}
\emph{the show must go on...}\\
\end{center}
\vspace{0.5cm}

Fourier's theory and its refinements that followed are unquestionably a success. As far as
the theory of musical signals goes, they represent a great achievement in elucidating the nature
of sounds and, more recently, their transformations. However, in order for it to be more
generally applicable, it must be further developed. Certain types of musical signals 
are not so well described by the current state of affairs; for example, what is the
representation of a musical signal where changes occur, such as for instance,
a mix of notes appearing and disappearing, glissandos, noise-like sounds, etc?
That surely has a spectrum, but what is it? J. Ville puts the question very
elegantly, \\

\begin{quote}
Si nous consid\'{e}rons en effet un morceau contenant plusieurs
mesures (ce qui est le moins qu'on puisse demander) et qu'une note, \emph{la} par exemple, figure une
fois dans le morceau, l'analyse harmonique nous pr\'{e}sentera la fr\'{e}quence correspondante avec une
certaine amplitude et une certaine phase, sans localiser le la dans le temps. Or, il est \'{e}vident qu'au
cours du morceau il est des instants o\`{u} l'on n'entend pas le \emph{la}. La repr\'{e}sentation est n\'{e}anmoins
math\'{e}matiquement correcte, parce que les phases des notes voisines du \emph{la} sont agenc\'{e}es de mani\`{e}re
\`{a} d\'{e}truire cette note par interf\'{e}rence lorsqu'on ne l'entend pas et \`{a} la renforcer, \'{e}galement par
interf\'{e}rence, lorsqu'on l'entend; mais s'il y a dans cette conception une habilet\'{e} qui honore l'analyse
math\'{e}matique, il ne faut pas se dissimuler qu'il y a \'{e}galement une d\'{e}figuration de la r\'{e}alit\'{e}: en
effet, quand on n'entend pas le \emph{la}, la raison v\'{e}ritable est que le \emph{la} n'est pas 
\'{e}mis\endnote{\emph{If we consider one piece containing several measures (which is the minimum we can ask for)
and that one note, \emph{A} for instance, appears one time in the piece, the harmonic analysis will present to us the
corresponding frequency with a certain amplitude and a certain phase, without localising it in time. Nevertheless,
it is evident that during the piece there are instants when we do not hear the \emph{A}. The representation is however
mathematically correct, because the phases of the neighbouring notes to the \emph{A} are acting in a manner to 
cancel this note by interference, when we do not hear it, or reinforce, also by interference, when we do; but if
there is in this concept a quality that honours the mathematical analysis, we should not dissimulate that there is
equally a disfiguration of reality: effectively when we do not hear the \emph{A}, the real reason is that the \emph{A} was not
emitted.} [A.T.]}. \cite[p.64]{Ville}\\
\end{quote}

The difficulties we face do not necessarily arise as a failure of the theory,  but from a fact
of reality, to which we have alluded earlier. We may say that on one hand it is not possible to define frequency 
without reference to time, and on the other, it is not possible to define time without reference to frequency. This
is something that becomes very apparent, for instance, when we try to arrive at a notion of \emph{instantaneous 
frequency}, which the work on frequency modulation highlighted (cf. \cite{Carson}). Parallel to this, developments in Quantum
Physics raised similar questions. In fact, one of the first attempts at tackling the problem of a dual representation 
appeared in E. Wigner's 1932 work \cite{Wigner} in that field. A similar approach was independently developed by J. Ville in 1948 within the scope of musical signals, defining \cite[p.72]{Ville}

\begin{equation}
F(t, f) = \frac 1 {2\pi} \int \Psi^{*}\left(t - \frac \nu {4\pi}\right)\Psi\left(t + \frac \nu {4\pi}\right) e^{-j\nu f } dt, 
\end{equation}\\

\noindent where the time domain is represented by the complex function $\Psi(t)$\endnote{This is
an analytic signal, in the form $\Psi(t) = s(t) + i\sigma(t)$, where $s(t)$ and $\sigma(t)$ are in
\emph{quadrature}, that is $\sigma(t) = -\pi^{-1} \int \frac {s(\tau) d\tau} { \tau - t}$.}. As can be seen we now
have a function of both (radian) frequency $f$ and time $t$. 

From another angle, D. Gabor follows a route from De Morgan's function segmentation. He 
proposes the use of a Gaussian envelope as a means of focusing the analysis 
at certain points in time \cite[p.435]{Gabor},\\

\begin{equation}
\psi(t) = e^{-\alpha^2(t - t_0)^2}\operatorname{cis}(2\pi f_0 t + \phi), 
\end{equation}\\

\noindent and utilising a complex sinusoid representation\endnote{Where $\operatorname{cis}(2\pi f_0 t + \phi) =
e^{2\pi f_0 t + \phi}$, is also an analytic signal.}
instead of a real signal to facilitate the analysis.
This produces a spectral component which is also subject to an envelope, 
this time in frequency, as its dual,\\

\begin{equation}
\phi(f) = e^{-\left(\frac {\pi} {\alpha}\right)^2  (f - f_0)}\operatorname{cis}[-2\pi t_0(f - f_0) + \phi]
\end{equation}\\

The pair can be used then to represent what Gabor termed \emph{acoustical quanta} \cite{Gabor2}. With this he was
able to define exactly the trade-off between time and frequency precision in the analysis.
This approach led to the development of the short-time Fourier transform and various very successful 
time-frequency musical signal processing techniques based on it. Both Wigner-Ville and Gabor opened
a new chapter in the story of the Fourier theory, with a lasting impact, but that unfortunately
is beyond the scope of this paper.

\newpage

\printendnotes

\vspace{0.5cm}

\nocite{*}
\bibliographystyle{amsrefs}
\bibliography{fourier}

\end{document}